\documentclass[a4paper,review,11pt,authoryear]{elsarticle}

\usepackage{natbib}
\usepackage{amsfonts,amsmath,bbm,bm,xcolor,booktabs,hyperref,amsthm}
\usepackage[ruled]{algorithm2e}
\usepackage{geometry}
\usepackage{subfig}
\usepackage{setspace}
\let\code=\texttt
\let\proglang=\textsf
\usepackage{soul}
\hypersetup{
  colorlinks=true,
  linkcolor=blue,
  citecolor=blue,
  urlcolor=blue}

\newcommand{\bY}{\mathbf{Y}}
\newcommand{\bpi}{\bm{\pi}}

\theoremstyle{definition}
\newtheorem{definition}{Definition}[section]

\begin{document}

\begin{frontmatter}

  \title{Discrete forecast reconciliation}

  \author[label1]{Bohan Zhang}
  \address[label1]{School of Economics and Management, Beihang University, Beijing, China}
  \author[label2]{Anastasios Panagiotelis}

  \author[label1]{Yanfei Kang\corref{cor1}}
  \ead{yanfeikang@buaa.edu.cn}
  \cortext[cor1]{Corresponding author.}
  \address[label2]{The University of Sydney Business School, NSW 2006, Australia}

  \begin{abstract}

    This paper presents a formal framework and proposes algorithms to extend forecast reconciliation to discrete-valued data to extend forecast reconciliation to discrete-valued data, including low counts. A novel method is introduced based on recasting the optimisation of scoring rules as an assignment problem, which is solved using quadratic programming. The proposed framework produces coherent joint probabilistic forecasts for count hierarchical time series.
    Two discrete reconciliation algorithms are also proposed and compared against generalisations of the top-down and bottom-up approaches for count data. Two simulation experiments and two empirical examples are conducted to validate that the proposed reconciliation algorithms improve forecast accuracy. The empirical applications are forecasting criminal offences in Washington D.C. and product unit sales in the M5 dataset. Compared to benchmarks, the proposed framework shows superior performance in both simulations and empirical studies.

  \end{abstract}


\end{frontmatter}

%\clearpage
\newpage
% \linenumbers

\section{Introduction}

Non-negative time series with discrete values, particularly those with low counts, commonly arise in various fields.
Examples include intermittent demand in the retail industry (\citealp{kourentzesElucidateStructureIntermittent2021}), occurrences of ``black swan'' events (\citealp{nikolopoulosWeNeedTalk2020}) and incidents of violent crime within a specific city block.
Count hierarchical time series (HTS) naturally arise when decisions must be made across multiple cross-sectional or temporal aggregation levels. Over the past decades, hierarchical forecasting has attracted significant attention from the forecasting community, yielding many innovative approaches and applications (see \citealp{athanasopoulosForecastReconciliationReview2023} and references therein). However, most existing hierarchical approaches are intrinsically designed for continuous-valued time series and can not be directly applied to discrete-valued data. This paper aims to fill this gap by proposing a discrete reconciliation approach for low-count HTS.

The core objective of hierarchical forecasting is to produce \textit{coherent} forecasts,
in the sense that the point forecast for a parent series should equal the sum of the point forecasts for the associated child series\footnote{For a discussion of how this can be extended to time series with general linear constraints, see \cite{girolimetto2023point}.}. Historically, this was accomplished by producing forecasts at one level
and then aggregating or disaggregating them into other levels (\citealp{fliednerHierarchicalForecastingIssues2001}).
An alternative approach, also known as \textit{forecast reconciliation}, proposed by \cite{hyndmanOptimalCombinationForecasts2011} and further explored by \cite{wickramasuriyaOptimalForecastReconciliation2019}, \cite{panagiotelisForecastReconciliationGeometric2021}, and others, involves first generating \textit{base} forecasts for each time series,
then optimally adjusting or \textit{reconciling} these to produce coherent forecasts. In most cases, this is achieved by taking a base forecast $\hat{\bm{y}}$ and premultiplying it by a projection matrix $\mathbf{P}$ to yield reconciled forecasts $\tilde{\bm{y}}=\mathbf{P}\hat{\bm{y}}$, that are coherent by construction. The precise form of the projection depends on assumptions about the covariance matrix of forecast errors, for instance, assuming homoskedastic uncorrelated forecasts leads to the OLS reconciliation of \cite{hyndmanOptimalCombinationForecasts2011}, while plugging in a shrinkage estimator of the forecast error covariance leads to the MinT method of \cite{wickramasuriyaOptimalForecastReconciliation2019}. Reconciliation can also be interpreted as a forecast combination method (\citealp{hollymanUnderstandingForecastReconciliation2021}) that avoids the requirement for complex models that simultaneously capture hierarchical constraints, external information, and serial dependence. In addition to achieving coherence, state-of-art reconciliation approaches have been demonstrated to improve forecast accuracy.

In a decision-making setting, it is natural to produce distributional forecasts rather than solely point forecasts such as the mean (\citealp{gneitingProbabilisticForecasting2014}). In the continuous data case, reconciliation has been extended to probabilistic forecasting by \cite{panagiotelisProbabilisticForecastReconciliation2022} building upon earlier efforts by \cite{jeonProbabilisticForecastReconciliation2019} and \cite{bentaiebHierarchicalProbabilisticForecasting2020}. Key components of \cite{panagiotelisProbabilisticForecastReconciliation2022} include an understanding that coherent forecasts lie on a linear subspace, that a linear function can map incoherent forecasts to this subspace, and that the weights of this linear function can be found by optimising proper scoring rules using first-order methods such as gradient descent. All of these components break down in a discrete data setting, necessitating a novel framework for defining coherence and optimally reconciling incoherent forecasts. These are the objectives that we address in this paper.

Before introducing our novel approach, it is essential to highlight the limitations of seemingly simpler alternatives. The first is to simply apply existing forecast reconciliation methods for continuous data, such as OLS or MinT projections, to discrete data. This would almost certainly yield non-integer forecasts. While this may be acceptable for mean forecasts, since the expected value of a discrete random variable need not be discrete, it will not yield a probabilistic forecast with the correct support. This may impact decisions that depend on specific probabilities, for example, the probability that no stock is sold on any day in a given week.  A second, simple solution may be using existing reconciliation methods and then rounding non-integer forecasts. Unfortunately, coherence will not, in general, be preserved after rounding. For example, in a hierarchy where $X+Y=Z$, the forecasts $X=0.4$, $Y=0.4$ and $Z=0.8$ are coherent, but rounding these values to the nearest integers leads to incoherence. As such, it makes more sense to develop a hierarchical forecasting approach that handles discrete distributions directly.

There have been other attempts to address count HTS forecasting.
\cite{coraniProbabilisticReconciliationCount2022} propose a novel reconciliation approach that conditions base probabilistic forecasts of the most disaggregated series on base forecasts of aggregated series.
The reconciled forecasts are derived by generalising Bayes’ rule and Monte Carlo sampling.
\cite{zambonEfficientProbabilisticReconciliation2022} further extend this idea to accommodate both count time series and real-valued time series. The main difference between this work and the methods we propose is a distinction between `conditioning' and `mapping'. Both \cite{zambonEfficientProbabilisticReconciliation2022}, and our own work proposed here, take an incoherent multivariate discrete distribution as an input.  \cite{zambonEfficientProbabilisticReconciliation2022} take the reconciled distribution to be a suitably normalised `slice' of the incoherent distribution along the domain where coherence holds. In contrast, what we propose in this paper trains a mapping from the domain where forecasts are incoherent to the domain where they are coherent. We argue that by training this `mapping', we are able to correct for model misspecification in the base forecast, including a proper accounting of dependence in the hierarchy. This is particularly appealing since for practical reasons, the input multivariate base models usually assume independence as multivariate discrete time series models remain challenging.

In this paper,  we first introduce the notion of ``coherence'' for hierarchical counts,
bearing a resemblance to the probabilistic coherence proposed by \cite{panagiotelisProbabilisticForecastReconciliation2022}.
Second, we utilise these concepts to establish a formal reconciliation framework for count HTSs, which reconciles the forecasts by assigning probability from incoherent to coherent domain points.
Third, we adopt the Brier score as a metric for evaluating forecasts and present a reconciliation algorithm that optimises this metric.
We also demonstrate how the choice of Brier score leads to an algorithm that can be solved through quadratic programming. To speed up computation, we also develop a second stepwise algorithm.
Fourth, two simulation experiments are performed to verify the applicability in both temporal and cross-sectional settings.
Finally, we conduct two empirical experiments using real data. The first experiment analyses a crime dataset with a temporal hierarchy, and the other employs cross-sectional sales data in the M5 dataset.

The remainder of this paper is organised as follows.
Section \ref{sec:coherence} presents notation and definitions of coherence and reconciliation for count HTSs.
Section \ref{sec:method} details the proposed discrete reconciliation framework based on optimising the Brier score through Quadratic programming. The stepwise version of the algorithm is also introduced in this section.
Section \ref{sec:simulation} performs two simulation experiments in cross-sectional and temporal settings, and two empirical experiments are conducted in Section \ref{sec:application}.
Then, Section \ref{sec:discussion} presents discussions and thoughts on future research. 
Finally, Section \ref{sec:conclusion} concludes the paper.
Data and code for reproducing the results in this paper are available at https://github.com/AngelPone/paper-dfr.

\section{Coherence of probabilistic hierarchical count forecasts}

\label{sec:coherence}

% \subsection{Preliminaries}

Consider an $n$-vector $\bY=\left(Y_1,Y_2,\ldots,Y_n\right)'$ of discrete random variables.
We partition $\bY$ so that the first $m$ elements are \textit{basis} variables and the remaining $(n-m)$ elements are \textit{determined} variables found as some linear combination of the basis variables. Usually, the basis variables are the bottom-level or most disaggregated data, while the determined variables are their aggregates.
Each element of $\bY$ has a finite domain given by $\mathcal{D}(Y_i)=\left\{0, 1,2,3,\dots,D_i\right\}$, where $i = 1, 2, \dots, n$.

\subsection{Discrete coherence}\label{sec:domains}

Defining coherence in the discrete setting first requires definitions of three sets upon which predictive distributions can be defined. The \textit{complete domain} of $\bY$ is given by
\[
\hat{\mathcal D}(\bY)=\prod\limits_{i=1}^n\hat{\mathcal D}(Y_i)=\left\{0, 1,2,\dots,D_1\right\}\times\left\{0,1,2,\dots,D_2\right\}\times\dots\times\left\{0,1,2,\dots,D_n\right\},
\]
where products are Cartesian products taken over sets, i.e. any possible vector where each element corresponds to a possible discrete value of a single variable.
The cardinality of the complete domain is $|\hat{\mathcal D}(\bY)|=\prod\limits_{i=1}^{n} (D_i+1)$, which is denoted by $q$.
The complete domain is analogous to $\mathbb{R}^n$ in the continuous case.

The \textit{coherent domain} of $\bY$, denoted as $\tilde{\mathcal D}(\bY)$, is given by a subset of $\hat{\mathcal D}(\bY)$, for which aggregation constraints hold.
It has cardinality $|\tilde{\mathcal D}(\bY)|=\prod\limits_{i=1}^{m} (D_i+1)$, which we denote as $r$.
The coherent domain is analogous to the coherent subspace $\mathfrak{s}$ in the continuous case (\citealp{panagiotelisProbabilisticForecastReconciliation2022}). The \textit{incoherent domain} $\bar{\mathcal D}(\bY)$ is defined as the set difference between the complete domain and incoherent domain, i.e. the set of points for which the aggregation constraints do not hold.

  \subsubsection*{\textbf{Example}}
  \label{sec:example}

  Let $Y_1$ and $Y_2$ be binary variables and $Y_3=Y_1+Y_2$. In this case, the domain of each variable is
  \[
    \mathcal{D}(Y_1)=\left\{0,1\right\},\quad
    \mathcal{D}(Y_2)=\left\{0,1\right\},\quad
    \mathcal{D}(Y_3)=\left\{0,1,2\right\}.
  \]
  The complete domain is
  \begin{equation}
  \begin{aligned}
  \hat{\mathcal D}(\bY)=&\left\{\mathbf{(0,0,0)'},(0,1,0)',(1,0,0)',(1,1,0)',\right.\\
  &\left.(0,0,1)',\mathbf{(0,1,1)'},\mathbf{(1,0,1)'},(1,1,1)',\right.\\
  &\left.(0,0,2)',(0,1,2)',(1,0,2)',\mathbf{(1,1,2)'}\right\}\,,
  \end{aligned}
  \label{eq:incoherent}
  \end{equation}
  and the coherent domain consists of those points for which $y_1+y_2=y_3$, highlighted in \textbf{bold} in Equation~\eqref{eq:incoherent}. Thus, the coherent domain is
  \[
      \tilde{\mathcal D}(\bY)=\left\{(0,0,0)',(0,1,1)',(1,0,1)',(1,1,2)'\right\}\,,
  \]
  while the incoherent domain $\bar{\mathcal D}(\bY)$ is the set of points for which the aggregation constraints do not hold, i.e.
    \[
  \bar{\mathcal D}(\bY)=\left\{(0,1,0)',(1,0,0)',(1,1,0)',(0,0,1)',
  (1,1,1)',(0,0,2)',(0,1,2)',(1,0,2)'
  \right\}\,,
  \]
  
  \begin{definition}[Discrete coherence]
  
  A discrete coherent distribution has the property of assigning zero probability to all coherent points, i.e. $Pr(\mathbf{Y}=\bm{y})=0, \forall \bm{y}\in \bar{\mathcal D}(\bY)$. Any distribution not meeting this condition is an incoherent distribution.

  \end{definition}

  Incoherent distributions arise in practice when probabilistic forecasts are generated independently for each variable, and the joint forecast is then constructed assuming independence. For instance, in the simple three-variable example provided, if  $Pr(Y_1=0)=0.2$,~$Pr(Y_2=1)=0.1$,~$Pr(Y_3=0)=0.05$, then under independence $Pr(Y_1=0,Y_2=1,Y_3=0)=0.2\times0.1\times0.05=0.001$, which assigns non-zero probability to an incoherent point.

  \subsection{Notation for Discrete Forecast Reconciliation}

  \label{sec:coherent_df}

  Dealing with an incoherent discrete distribution and a coherent discrete distribution, each with a different domain of support, requires the definition of some notational conventions. Consider an $h$-step ahead (incoherent) base forecast for $\bY$ at time $t$. This has support on $q$ points and can thus be described by $q$-dimensional probability vector $\hat{\bpi}^{t+h|t}$, where each element corresponds to the probability of one discrete point in the complete domain. A one-to one mapping function $\hat{\mathcal{H}}:\{1,2,\dots,q\}\rightarrow\hat{\mathcal{D}}(\bY)$, is used to match an index of the elements of $\hat{\bpi}^{t+h|t}$ to a configuration of values that $\bY$ can take. Two notational conventions are used for the elements of $\hat{\bpi}^{t+h|t}$.
  First, $\hat{\pi}_j^{t+h|t}$ denotes the $j^{th}$ element of $\hat{\bpi}^{t+h|t}$;
  second, $\hat{\pi}_{(y_1 y_2 \dots y_n)}^{t+h|t}$ denotes a specific element that corresponds to the forecast probability that $\bY$ takes on a value $(y_1,y_2,\dots,y_n)'$. 
  Using the small example in Section~\ref{sec:domains}, $\hat{\pi}_1^{t+h|t}=\hat{\pi}_{(000)}^{t+h|t}$, $\hat{\pi}_2^{t+h|t}=\hat{\pi}_{(010)}^{t+h|t}$, etc., and $\hat{\mathcal{H}}(1)=(0,0,0)'$, $\hat{\mathcal{H}}(2)=(0,1,0)'$, etc. Since this vector representation uniquely characterises a probability mass function, `probability vectors', `pmfs' and `distribution' will be used interchangeably throughout the rest of the paper.

  On the other hand, a coherent probabilistic forecast can be defined using an $r$-vector $\tilde{\bpi}^{t+h|t}$ where each element corresponds to the probability assigned to a point in the coherent domain.
  The notation $\tilde{\pi}_k^{t+h|t}$ represents the $k^{th}$ element of $\tilde{\bpi}^{t+h|t}$ and the notation $\tilde{\pi}_{(y_1 y_2 \dots y_n)}^{t+h|t}$ denotes a specific element of this vector that corresponds to the forecast probability that $\bY$ takes on a coherent value $(y_1,y_2,\dots,y_n)'$.
  The analogue to $\hat{\mathcal{H}}(j)$ is a function  $\tilde{\mathcal{H}}:\{1,2,\dots,r\}\rightarrow\tilde{\mathcal{D}}(\bY)$, which maps each index $k$ to a coherent configuration of values that $\bY$ can take.

  \subsubsection*{\textbf{Example}}

  Consider the earlier example in Section~\ref{sec:domains}, with binary $y_1$ and $y_2$ and $y_1+y_2=y_3$. The incoherent probabilistic forecast is given by
  \[
    \hat{\bpi}^{t+h|t}= \left[
      \hat{\pi}^{t+h|t}_{(000)}, ~
       \hat{\pi}^{t+h|t}_{(010)},~
       \hat{\pi}^{t+h|t}_{(100)},~
       \hat{\pi}^{t+h|t}_{(110)},~
       \hat{\pi}^{t+h|t}_{(001)},~
       \hat{\pi}^{t+h|t}_{(011)},~
       \hat{\pi}^{t+h|t}_{(101)},~
       \hat{\pi}^{t+h|t}_{(111)},~
       \hat{\pi}^{t+h|t}_{(002)},~
       \hat{\pi}^{t+h|t}_{(012)},~
       \hat{\pi}^{t+h|t}_{(102)},~
       \hat{\pi}^{t+h|t}_{(112)}
       \right]',
  \]
  where the notation $\hat{\pi}^{t+h|t}_{1}$ can be used instead of $\hat{\pi}^{t+h|t}_{(000)}$, and $\hat{\pi}^{t+h|t}_{2}$ can be used instead of $\hat{\pi}^{t+h|t}_{(010)}$, etc. Also, the function $\hat{\mathcal{H}}$ is defined so that $\hat{\mathcal{H}}(1)=(0,0,0)'$, $\hat{\mathcal{H}}(2)=(0,1,0)'$, etc.

  The coherent probabilistic forecast is given by
  \[
  \tilde{\bpi}^{t+h|t}=\left[
  \tilde{\pi}^{t+h|t}_{(000)},
  \tilde{\pi}^{t+h|t}_{(011)},
  \tilde{\pi}^{t+h|t}_{(101)},
  \tilde{\pi}^{t+h|t}_{(112)}
  \right]',\]
  where the notation $\tilde{\pi}^{t+h|t}_{1}$ can be used instead of $\tilde{\pi}^{t+h|t}_{(000)}$, and $\tilde{\pi}^{t+h|t}_{2}$ can be used instead of $\tilde{\pi}^{t+h|t}_{(011)}$, etc. Also, the function $\tilde{\mathcal{H}}$ is defined so that $\tilde{\mathcal{H}}(1)=(0,0,0)'$, $\tilde{\mathcal{H}}(2)=(0,1,1)'$, etc. Note that the ordering of the probabilities and the functions $\hat{\mathcal{H}}$ and $\tilde{\mathcal{H}}$ are not unique, which does not affect the proposed algorithms.

  \subsection{Discrete forecast reconciliation}\label{sec:discrec}
  
   \begin{definition}[Discrete reconciliation]
  	
  	Let $\mathcal{C}^{b}$ denote a $b$-dimensional unit simplex and $\psi:\mathcal{C}^{q-1} \rightarrow \mathcal{C}^{q-1}$.	Then a reconciled discrete forecast is given by 
  	\[
  	  \tilde{\bpi} = \psi(\hat{\bpi})\,,
  	\]
    where the superscript $t+h|t$ dropped for convenience.
   \end{definition}
    
   Loosely speaking, the discrete reconciliation framework constructs the coherent distribution by assigning mass from incoherent points to coherent points in the predictive distribution. This framework is analogous to that described in \cite{panagiotelisProbabilisticForecastReconciliation2022}, where an incoherent probability measure on $\mathbb{R}^n$ is mapped onto the coherent subspace $\mathfrak{s}$.

   In this paper, we focus on the linear reconciliation function given by
    \begin{equation}
	    \label{eq:framework}
	    \tilde{\bpi}=\bm{A}\hat{\bpi},
    \end{equation}
   where $\tilde{\bpi}$ is obtained by multiplying the $r \times q$ matrix of reconciliation weights $\bm{A}$ with the incoherent probability vector $\hat{\bpi}$. Letting $a_{kj}$ be the element in row $k$ and column $j$ of $\bm{A}$, this is equivalent to
    \[
      \tilde{\pi}_k=\sum\limits_{j=1}^q a_{kj}\hat{{\pi}}_j
    \]
    for all $k = 1, 2, \dots, r$  and  $j = 1, 2, \dots, q$.
   Each $a_{kj}$ represents how much probability is shifted from the possibly incoherent point corresponding to element $j$ in $\hat{\bpi}$ to a coherent point corresponding to element $k$ in $\tilde{\bpi}$ (or from $\hat{\mathcal{H}}(j)$ to point $\tilde{\mathcal{H}}(k)$). Note that the matrix $\mathbf{A}$ is not a projection matrix as used for continuous data, and $a_{kj}$ must meet the following constraints
   \begin{align*}
	 0\leq a_{kj} \leq 1 , ~ \forall k, j, ~ \textrm{and} ~
	\sum\limits_{k=1}^r a_{kj} = 1 , ~ \forall j.
    \end{align*}
    The first constraint guarantees that the elements of $\tilde{\bpi}$ are between 0 and 1, while the second constraint guarantees that the elements of $\tilde{\bpi}$ sum to 1. Moreover, they imply that every point in the incoherent domain has its probability proportionally distributed among all points within the coherent domain. This process resembles the assignment problem commonly encountered in operational research.

\section{Method}
\label{sec:method}

This section outlines our proposed methods for optimal discrete reconciliation. Section~\ref{sec:algorithm} introduces a framework for training the reconciliation matrix in Section~\ref{sec:discrec} by optimising the Brier Score. To address the issue of dimensionality, we further propose a stepwise reconciliation algorithm that decomposes the hierarchy in Section~\ref{sec:algorithm2}. Additionally, Section~\ref{sec:bottomup} extends the classical bottom-up and top-down approaches to the discrete case and demonstrates how they can be incorporated into our framework.

    \subsection{Score optimal reconciliation}
    \label{sec:algorithm}

    Scoring rules $S(.,.)$ are commonly used to evaluate probabilistic forecasts and assign a numerical score based on the predictive distribution and the actual outcome.
    An essential property for scoring rules is to be strictly proper, i.e., $\text{E}_Q[S(Q, \mathbf{y})] \leq \text{E}_Q[S(P, \mathbf{y})]$ with equality if and only if $P=Q$, where $P$ is any predictive distribution, $Q$ is the true distribution and $\mathbf{y}$ is an outcome.
    Strictly proper scoring rules that can be used to evaluate discrete distributions include the Brier Score, spherical Score and logarithmic Score (see \citealp{gneitingStrictlyProperScoring2007} for more details).
    The Brier Score was initially proposed by \cite{brier1950verification}, and it has the following formulation:
    \[
      \text{BS}(\bpi, \mathbf{z}) = \sum_{k=1}^{O}(z_k - \pi_k)^2,
    \] where $\bpi$ is defined similarly as in Section~\ref{sec:coherent_df},
    $k\in \{1,\dots,O\}$ is the index that can be mapped into the potential outcomes of an event through link function $\mathcal{H}$, and $\mathbf{z}$ is a vector with $z_k = 1$ if $\mathcal{H}(k) = \mathbf{y}$ and $0$ otherwise.
    We choose the Brier score instead of other strictly proper scoring rules since it leads to an objective function that easily can be solved through quadratic programming.

\begin{figure}
    \centering
    \includegraphics[width=0.5\textwidth]{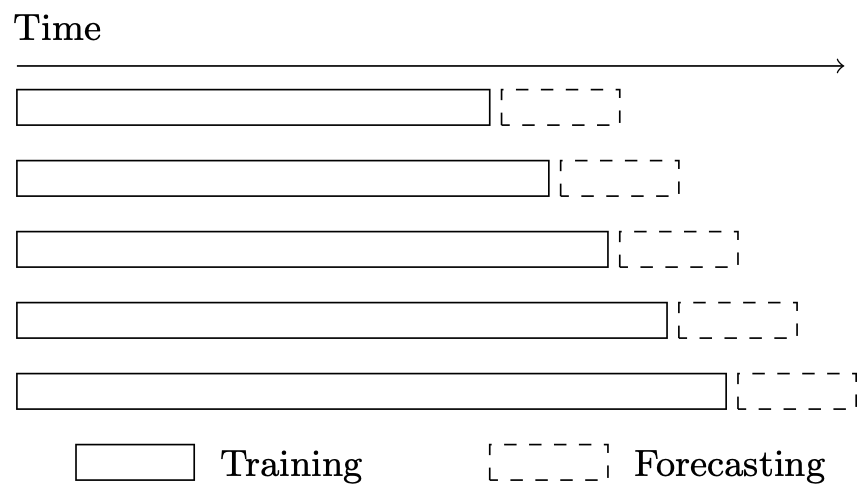}
    \caption{\label{fig:rollingwindow} Diagram of expanding window strategy.}
    \end{figure}

    Assume that $\hat{\bpi}^{t+h|t}$ are found for $t\in\mathcal{T}_{\textrm{window}}$, where $\mathcal{T}_{\textrm{window}}$ is a rolling or expanding window (\citealp{hyndmanForecastingPrinciplesPractice2021}).
    Figure~\ref{fig:rollingwindow} depicts the behaviour of the expanding window: training data always begins at the same origin, one observation is added to the training data for each window, and forecasts are made $h$ steps ahead. This process continues until the last observation is included in the forecast horizon.
    Letting $\mathbf{z}^{t+h}$ be an $r$-vector with element $z_k^{t+h}=1$ if $\tilde{\mathcal{H}}(k)=\bm{y}^{t+h}$ and $0$ otherwise, where $\bm{y}^{t+h}$ is the actual realisation of $Y$ at time $t+h$, the Brier Score can be averaged over the expanding windows as
    \begin{align*}
    {\overline{BS}}=& \frac{1}{|\mathcal{T}_{\textrm{window}}|}\sum\limits_{\mathcal{T}_{\textrm{window}}}\left[(\mathbf{A}\hat\bpi^{t+h|t} - \mathbf{z}^{t+h})'(\mathbf{A}\hat\bpi^{t+h|t} - \mathbf{z}^{t+h})\right] \\
    =& \frac{1}{|\mathcal{T}_{\textrm{window}}|}\sum\limits_{\mathcal{T}_{\textrm{window}}}\left[\sum\limits_{k=1}^r\left(\tilde{\pi}_k^{t+h|t}-z^{t+h}_k\right)^2\right]\\
    =& \frac{1}{|\mathcal{T}_{\textrm{window}}|}\sum\limits_{\mathcal{T}_{\textrm{window}}}\left[\sum\limits_{k=1}^r\left(\sum\limits_{j=1}^q a_{kj}\hat{{\pi}}_j^{t+h|t}-z^{t+h}_k\right)^2\right]\,.
    \end{align*}
    This is a quadratic function of $a_{kj}$ with smaller values indicating a better coherent forecast.

    \subsubsection*{\textbf{Movement restriction}}
    While the Brier Score of the reconciled distribution can be minimised, this can be computationally costly as $q$ and $r$ grow. This can be mitigated by introducing further restrictions to the elements of $\mathbf{A}$, which we refer to as `movement' restrictions.  The intuition behind these restrictions is as follows. When assigning probability from incoherent points, probabilities should be assigned to coherent points that are \emph{nearby} in some sense. This idea is inspired by the successful use of projections in the reconciliation literature for continuous forecasts, where an incoherent forecast is mapped to the \emph{nearest} point on the coherent subspace. We define the \emph{cost} of moving probability from $\hat{\pi}_j\rightarrow\tilde{\pi}_k$ as
    \[
    c_{kj}=||\hat{\mathcal{H}}(j)-\tilde{\mathcal{H}}(k)||_1\,,
    \]
    where $||.||_1$ represents $\mathcal{L}_1$ norm. For example, in the three-variable scenario, the cost between $(0, 1, 0)'$ and $(0, 0, 0)$' is
    \[
    c_{12}=\left|\left|\begin{pmatrix}0\\1\\0\end{pmatrix}-\begin{pmatrix}0\\0\\0\end{pmatrix}\right|\right|_1=1\,.
    \]
    It should be noted that, unlike the continuous case, there may not be a unique nearest coherent point; for instance, $(0,1,0)'$ is equally distant from $(0,0,0)'$ and $(0,1,1)'$, which are both coherent.
    Based on the costs, we force $a_{kj}=0$ if
    \[
      c_{kj}>\underset{k'}{\min}\,c_{k'j}\,,
    \]
    which implies that probability can only be moved to one of the nearest points. In the three-variable example, probability can be moved from $(0,1,0)'$ to $(0,0,0)$ and $(0,1,1)$ but not to $(1,0,1)$ and $(1,1,2)$.
    Additionally, $a_{kj}=1$ for all $k,j$ such that $\hat{\mathcal{H}}(k)=\tilde{\mathcal{H}}(j)$.
    This restriction implies that all probability from a coherent point in the complete domain is assigned to the same point in the coherent domain.

    \subsubsection*{\textbf{Objective}}

    The final objective function is defined as follows:

    \[
    \underset{a_{kj}}{\min} \frac{1}{|\mathcal{T}_{\textrm{window}}|}\sum\limits_{\mathcal{T}_{\textrm{window}}}\left[\sum\limits_{k=1}^r\left(\sum\limits_{j=1}^q a_{kj}\hat{{\pi}}_j^{t+h|t}-z^{t+h}_k\right)^2\right]
    \]
    subject to
    \[
    \begin{aligned}
    &0\leq a_{kj}\leq 1,\forall j, k \quad \text{and} \quad
    \sum\limits_{k=1}^r a_{kj} = 1,~\forall j,\\
    & a_{kj} = 0 \quad \forall k,j: c_{kj}>\underset{k'}{\min}\,c_{k'j}\,.
    \end{aligned}
    \]
    This objective is a standard quadratic programming problem, which can be efficiently solved using quadratic solvers such as the Operator Splitting Solver~\citep[OSQP, ][]{stellatoOSQPOperatorSplitting2020}.

    While the proposed framework could accept any incoherent probabilistic forecast as input, for the remainder of the paper, we consider base distributions that are independent with margins derived from univariate forecasting methods for count time series. This approach is motivated by the wider availability of univariate time series models for count data compared to their multivariate counterparts. In this setting, the $j$-th element of the resulting probability vector is calculated as follows: \[
    \hat{\pi}_j = \hat{\pi}_{(y_1,y_2,\dots,y_n)} = \hat P_{1}(Y_1=y_1)\times\dots\times\hat P_{n}(Y_n=y_n).
    \] 
    
    Note that despite independence in the base forecast, reconciliation will induce dependence in the reconciled forecast via the training of the $\bm{A}$ matrix.
    
\begin{figure}
    \centering
    \includegraphics[width=\textwidth]{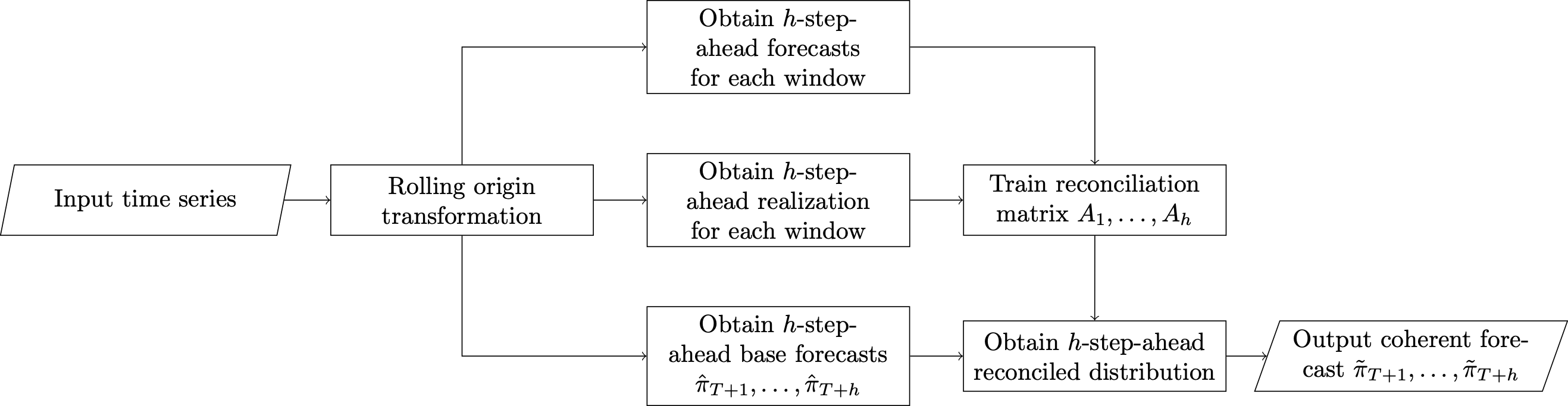}
    \caption{\label{fig:dfr}Flowchart for the DFR algorithm.}
    \end{figure}

    Combining all the elements considered above, we propose the Discrete Forecast Reconciliation (DFR) algorithm. The architecture of the proposed approach is illustrated in Figure~\ref{fig:dfr}. It consists of two stages: training and forecasting.   In the training stage, a time series of length $T$ is split into $|\mathcal{T}_{\text{window}}|$ windows.
    For each window, $h$-step-ahead base forecasts $\hat \bpi$ are generated using univariate forecasting tools and assuming independence and paired with corresponding realisations $\mathbf{z}$.
    Using these pairs, we find the reconciliation matrix $\mathbf{A}$ by optimising the objective function described in Section~\ref{sec:algorithm}.
    In the forecasting stage, the coherent joint distribution is obtained by multiplying the trained reconciliation matrix with the vector of probabilities from the base incoherent distribution.
    Multi-step ahead forecasts are obtained by building a separate reconciliation model for each step.

    \subsection{Stepwise discrete reconciliation}
    \label{sec:algorithm2}

    Although the so-called movement restriction reduces the number of free parameters in the quadratic program, as the number of variables increases, the number of elements in $\mathbf{A}$ grows exponentially.
    As a result, it is difficult to handle even a hierarchy of moderate dimension using the DFR algorithm alone.
    To address this issue, we propose a Stepwise Discrete Forecast Reconciliation (SDFR) algorithm that can overcome computational issues associated with high dimensionality.
    Instead of reconciling forecasts of all series at once,  SDFR
    decomposes a high-dimensional hierarchy into multiple two-level hierarchies each with three nodes. These sub-hierarchies are reconciled step by step using the DFR algorithm shown in Section~\ref{sec:algorithm}.
    The stepwise procedure requires adjustments to the reconciled forecasts of the sub-hierarchies at each step. The training stage of SDFR is identical to DFR, while the forecasting stage is shown in Algorithm~\ref{alg:stepwise}.
    \begin{algorithm}[ht]
    \caption{\label{alg:stepwise}\textbf{S}tepwise \textbf{D}iscrete \textbf{F}orecast \textbf{R}econciliation (\textbf{SDFR})}
    \SetKwFunction{reconcile}{DFR$_i$}
    \SetKwFunction{bu}{BottomUp}
    \SetKwFunction{adjust}{Adjust}
    \SetKwFunction{construct}{ConstructJointDist}
    \SetKwInOut{Input}{Input}
    \SetKwInOut{Output}{Output}
    \Input{$\hat{\pi}_0,\dots,\hat{\pi}_m$}
    \For {$i=1,\dots,m-1$}{
      $\hat{\pi}_{\mathbf{S}_{m-i}} \leftarrow$ \bu($\hat\pi_{i+1},\dots,\hat\pi_m$)\;
      \uIf{ i = 1}{
        $\hat{\pi}_{\mathbf{S}_{m-i+1}} \leftarrow \hat\pi_0$ \;
      }
      \Else{$\hat{\pi}_{\mathbf{S}_{m-i+1}} \leftarrow \sum_{\mathbf{S}_{m-i+2}, y_{i-1}}\tilde{\bpi}(\mathbf{S}_{m-i+2}, y_{i-1}, \mathbf{S}_{m-i+1})$\;
      }

      $\tilde{\pi}(\mathbf{S}_{m-i+1}, y_i, \mathbf{S}_{m-i}) \leftarrow$ \reconcile{$\hat{\pi}_{\mathbf{S}_{m-i + 1}}, \hat\pi_{i}, \hat\pi_{\mathbf{S}_{m-i}}$}
    }

    \For {$i=2,\dots,m-1$} {
      $\tilde\pi^{1}_{\mathbf{S}_{m-i+1}} \leftarrow \sum_{\bY_{i-1}}\tilde{\bpi}(\bY_{i-1}, \mathbf{S}_{m-i+1})$ \;
      $\tilde\pi^{2}_{\mathbf{S}_{m-i+1}} \leftarrow \sum_{y_i,\mathbf{S}_{m-i}}\tilde{\bpi}(\mathbf{S}_{m-i+1}, y_i, \mathbf{S}_{m-i})$ \;
      $\tilde\pi'_{\mathbf{S}_{m-i+1}} \leftarrow \frac{1}{2} (\tilde\pi^{1}_{\mathbf{S}_{m-i+1}} + \tilde\pi^{2}_{\mathbf{S}_{m-i+1}}$) \;
      $\tilde{\bpi}'(\bY_{i-1}, \mathbf{S}_{m-i+1}) \leftarrow$ \adjust($\tilde{\bpi}(\bY_{i-1}, \mathbf{S}_{m-i+1}), \tilde{\pi}'_{\mathbf{S}_{m-i+1}})$ \;
      $\tilde{\bpi}'(\mathbf{S}_{m-i+1}, y_i, \mathbf{S}_{m-i}) \leftarrow$ \adjust($\tilde{\bpi}(\mathbf{S}_{m-i+1}, \mathbf{S}_{m-i+1}), y_i, \tilde{\pi}'_{\mathbf{S}_{m-i+1}})$ \;
      $\tilde{\bpi}(\bY_i, \mathbf{S}_{m-i}) \leftarrow$ \construct($\tilde{\bpi}'(\bY_{i-1}, \mathbf{S}_{m-i+1}), \tilde{\bpi}'(\mathbf{S}_{m-i+1}, y_i, \mathbf{S}_{m-i}) $)\;
    }
    \Output{$\tilde \bpi(\bY_m)$}

    \end{algorithm}

  Taking a hierarchy with one total series and $m~(m>2)$ bottom series as an example, Algorithm \ref{alg:stepwise} shows how independently generated base forecasts can be reconciled into coherent joint forecasts step-by-step when $m$ is large.
  Denote the total series as $y_0$ and bottom-level series as $y_1, \dots, y_m$.
  Denote the vector of first $i+1$ variables as $\mathbf{Y}_i$ and the sum of last $j$ variables as $\mathbf{S}_j$, i.e.,
  \[
    \bY_i = (y_0, \dots, y_i)', \quad \mathbf{S}_j = \sum_{l=m-j+1}^{m} y_l.
  \]
  In general, the hierarchy is split into $m-1$ three-node hierarchies.
  For each hierarchy $i$, one bottom series (the left node) corresponds to the node $i+1$ in the original hierarchy, i.e., $y_{i}$. The other bottom series (the right node) corresponds to $\mathbf{S}_{m-i}$, which is the sum of the remaining $m-i$ series; thus making $\mathbf{S}_{m-i+1}$ their total node.
  In the training stage, the reconciliation model \code{DFR}$_i$ is trained for this hierarchy.
  Base forecasts of the left node are obtained from input, while base forecasts of the right node are obtained using a bottom-up approach that is discussed in Section~\ref{sec:bottomup}.
  The base forecasts for the total node are derived from the marginal distribution of that same node, which is derived from the coherent distribution obtained in the previous step.
  For cross-sectional hierarchies, base forecasts can be obtained for the right and total node by forecasting $\mathbf{S}_{j}, j=2,\dots,m-1$ directly.
  However, in temporal hierarchies, forecasting $\mathbf{S}_{m-i+1}$ requires forecasting temporal aggregation of partial time periods, introducing non-integer frequency problems and making it more challenging to capture time series dynamics.
  Our approach instead offers a simple implementation with no extra expert intervention or modelling and can be applied to both cross-sectional and temporal hierarchies.

  During the forecasting stage, we first pass the base forecasts stepwise into these models to obtain $m-1$ coherent forecasts.
  Adjacent hierarchies share the same node (i.e., $\mathbf{S}_{m-i+1}$), but their marginal forecasts for the shared node (i.e., $\tilde\pi^{2}_{\mathbf{S}_{m-i+1}}$ and $\tilde\pi^{1}_{\mathbf{S}_{m-i+1}}$) are not identical since reconciliation generally changes input forecasts.
  We average these two marginal forecasts and pass their average into the \code{Adjust} algorithm. This algorithm adjusts the joint distributions from two adjacent steps to ensure the marginal distribution of the shared node is equivalent to the average across the two steps.
  The two adjusted distributions are then passed into the \code{ConstructJointDist} algorithm to construct a new joint distribution.
  For brevity, the \code{Adjust} and \code{ConstructJointDist} algorithms are shown in \ref{appendix:adjust}.
  Figure~\ref{fig:sdfr} presents an example explaining how the SDFR algorithm works, where a four-node hierarchy is split into two three-node hierarchies. The DFR algorithm is applied stepwise from left to right to each three-node hierarchy.

  \begin{figure}
    \centering
    \includegraphics[width=0.5\textwidth]{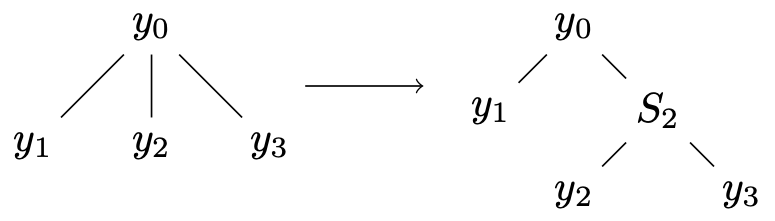}
    \caption{\label{fig:sdfr}Example procedure of the SDFR algorithm.}
  \end{figure}

  Note that in our proposed algorithm, the result is sensitive to the order in which the bottom-level series are combined. Consequently, we average results over different random orders of the bottom-level series; this reduces forecast uncertainty introduced by the \code{Adjust} and \code{ConstructJointDist} algorithms.
  When handling a hierarchy with more aggregation levels, we can repeat the above procedure for each aggregated series either in a bottom-up or top-down manner. This means we reconcile the base forecasts of one level and use the reconciled forecasts as base child forecasts (bottom-up) or base parent forecasts (top-down) to reconcile the next level.

    \subsection{Probabilistic extensions of bottom-up and top-down methods for count series}

    Having extended the framework of forecast reconciliation to discrete data, we can also define discrete analogues to the traditional top-down and bottom-up reconciliation methods as special cases. We refer to these algorithms as Discrete Bottom-Up (DBU) and Discrete Top-Down (DTD) methods. These will be used as benchmarks in our empirical examples. These methods only require forecasts at a single level.
    
    \subsubsection*{\textbf{Discrete bottom-up}}
    \label{sec:bottomup}

    The discrete bottom-up method constructs a coherent distribution by assuming independent bottom-level forecasts.
    This method is a special case of the algorithm described in Section~\ref{sec:algorithm} except that the base forecasts of aggregated series are ignored. A full vector of incoherent probabilities is formed by assuming independence (i.e. in the three-variables example, $\tilde{\pi}_{(000)} = Pr(Y_1=0)\times Pr(Y_2=0)$), and \[
    \mathbf{A} = [\mathbf{I}_4, \quad \mathbf{I}_4, \quad \mathbf{I}_4 ].
    \]
    The mean forecast obtained from this coherent distribution is identical to that obtained by directly aggregating mean forecasts of bottom-level series. Alternative bottom-up methods that learn the dependence structure between bottom-level time series in the spirit of \citealp{bentaiebHierarchicalProbabilisticForecasting2020}) could also be derived. However, this lies beyond the scope of our current work.
    
    \subsubsection*{\textbf{Discrete top-down}}

    The discrete top-down method extends the traditional top-down method. A challenge in doing so is to apportion probability to points that lead to the same value of the top-level aggregate. Consider the simple example from Section~\ref{sec:coherence}. While $Pr(Y_3=0)$ and $Pr(Y_3=2)$ should be apportioned to the points $(0,0,0)$ and $(1,1,2)$ respectively, the probability $Pr(Y_3=1)$ could be assigned to either $(0,1,1)$ or $(1,0,1)$. This probability can be split according to historical proportions. For example, if there are $40$ $(0, 1, 1)$ points and $60$ $(1, 0, 1) $ points observed and if  $Pr(Y_3=1) = 0.1$, then $\tilde \pi_{(011)} = 0.1\times 0.4$ and $\tilde \pi_{(101)} = 0.1\times 0.6$.
    This method can also be considered as a special case of Equation \eqref{eq:framework}, where a full vector of probabilities on the complete domain is found by apportioning the marginal probabilities according to historical proportions and
    \[
    \mathbf{A} = \left[\begin{matrix}
      1 & 1 & 1 & 1 & 0 & 0 & 0 & 0 & 0 & 0 & 0 & 0 \\
      0 & 0 & 0 & 0 & 0.4 & 0.4 & 0.4 & 0.4 & 0 & 0 & 0 & 0 \\
      0 & 0 & 0 & 0 & 0.6 & 0.6 & 0.6 & 0.6 & 0 & 0 & 0 & 0 \\
      0 & 0 & 0 & 0 & 0 & 0 & 0 & 0 & 1 & 1 & 1 & 1
    \end{matrix}\right].
    \]

\section{Simulation}
\label{sec:simulation}

To showcase the effectiveness of our proposed framework, we conduct two simulation experiments in distinct contexts: one is a cross-sectional hierarchy the other a temporal hierarchy.

  \subsection{Cross-sectional hierarchy}
  \label{sec:cross-sectional_simu}
    \subsubsection{Simulation setup}
    This subsection considers the three-variable hierarchy depicted in Section~\ref{sec:example}. The binary count time series at the bottom level have a data generating process based on an underlying latent $2\times 1$ vector $\mathbf{s}_t$ that follows a bivariate VAR(1) process
    \[\mathbf{s}_t = \mathbf{\Phi}\mathbf{s}_{t-1}+\boldsymbol{\eta}_t,\]
    where
    \[
      \mathbf{\Phi} = \left[\begin{matrix}
        \alpha & 0 \\
        0 & \beta
      \end{matrix}\right], ~ \boldsymbol{\eta} \sim \mathcal{N}\left(\mathbf{0}, \left[\begin{matrix}
        0.1 & 0.05 \\
        0.05 & 0.1
      \end{matrix}\right]\right), ~ \mathbf{s}_{0} = \left[
        \begin{matrix}0 \\ 0\end{matrix}
      \right],
    \]
    Observations $Y_{it}$ for $i=1,2$, are set to 1 when $s_{it}>0$ and 0 otherwise and $Y_{3t}=Y_{1t}+Y_{2t}$. The parameters $\alpha$ and $\beta$ are uniformly generated from $[0.4, ~ 0.5]$ and $[0.3, ~ 0.5]$ respectively, ensuring stationarity. Correlation in the error term $\boldsymbol{\eta}$ implies a positive correlation between the observed bottom-level series.

    Using the data-generating process described above, $480$ observations were simulated. Using an expanding window strategy with training samples of $150$ observations, we obtain $330$ pairs of one-step ahead forecasts and realisations. Using the first $300$ of these pairs, a matrix $\mathbf{A}$ is trained according to the DFR algorithm. The final $30$ observations are used to evaluate DFR against other methods. The entire process is replicated 1000 times.

    The base probabilistic forecasts are obtained independently using the binomial AR(1) model~(\citealp{weissParameterEstimationBinomial2013}), which is suitable for count time series with a finite range. This model is misspecified, reflecting a realistic scenario where the true DGP is never known in practice. As well as evaluating base forecasts, three reconciliation methods are considered: top-down, bottom-up and the DFR method. We also consider the empirical distribution of $\mathbf{Y}$, which does not account for serial dependence, but is a robust and commonly used benchmark for low-count time series forecasting. To evaluate forecasts, we compute the Brier scores for the marginal distribution of each variable, and for the joint distribution. The Brier scores are averaged over the 1000 replications and 30 out-of-sample evaluation points.

    \subsubsection{Simulation results}
    Table~\ref{tab:sim_crosssectional_res_dist} summarises the average Brier scores for all five methods, with the best performing method shown in bold for each row. The Brier scores for the base forecasts are low for $Y_1$ and $Y_2$ when compared to $Y_3$, suggesting a higher degree of misspecification for the top-level series when using the binomial AR(1). The base forecasts have lower scores than the empirical distribution when evaluated marginally, which is unsurprising since the empirical benchmark does not account for serial dependence. The Brier score for the joint base distribution, however, is high, which is a consequence of incoherence. After performing reconciliation with the DFR method, there is a large improvement in the Brier score for $Y_3$, which offsets the slight deterioration in the Brier score for the bottom-level series. There is also a substantial improvement in the Brier score for the joint distribution.
    
\begin{figure}
	\centering
	\caption{Average ranks and 95\% confidence intervals for the four approaches in the cross-sectional simulation. The average ranks of each approach in terms of Brier scores are shown to the right of its name.}
	\includegraphics[width=\textwidth]{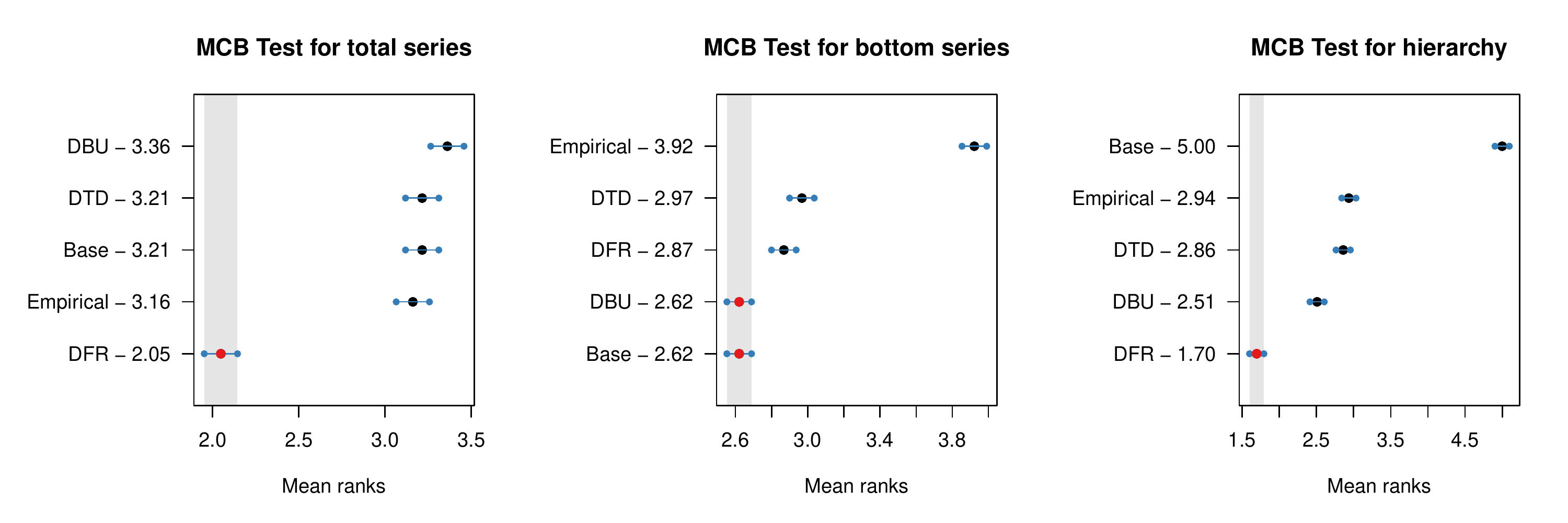}
    \label{fig:mcb_crosssectional} 
    \end{figure}

    To assess whether the differences in Brier score reported in Table~\ref{tab:sim_crosssectional_res_dist} are statistically significant, Multiple Comparisons from the Best (MCB) tests (\citealp{koningM3CompetitionStatistical2005}) were performed at a 95\% confidence level.
    This testing approach is rank-based and, therefore, does not make any assumptions about the distribution of Brier scores. These are summarised in Figure~\ref{fig:mcb_crosssectional}. If the bar corresponding to a method does not overlap with the grey intervals shown around the best method, this indicates statistically significant differences between methods.
    Figure~\ref{fig:mcb_crosssectional} shows that the DFR method performs significantly better than all other approaches for the total series and the hierarchy (left and right panels). The middle panel of Figure~\ref{fig:mcb_crosssectional} is a composite result for both bottom-level series. While DFR leads to a slight deterioration in the Brier score at the bottom level compared to the base forecasts, this is not statistically significant.

    \begin{table}
      \centering
      \caption{\label{tab:sim_crosssectional_res_dist} Probabilistic forecast performance summary for cross-sectional setting with Brier Score ($\times 10^{-2}$). Row-wise minimum values are displayed in \textbf{bold}.}
      \begin{tabular}{lccccc}
      \toprule
      % \multicolumn{5}{c}{Brier Score ($\times 10^{-2}$)}\\
      ~ & Base & DBU & DTD & DFR & Empirical \\ \midrule
      $Y_1$ & \textbf{46.52} & \textbf{46.52} & 47.25 & 46.78 & 50.41 \\ 
      $Y_2$ & \textbf{46.87} & \textbf{46.87} & 47.79 & 47.32 & 50.43 \\ 
      $Y_3$ & 67.07 & 67.13 & 67.07 & \textbf{64.57} & 67.15 \\ 
      $\mathbf{Y}$ & 88.65 & 71.23 & 72.00 & \textbf{69.60} & 72.9 \\ 
      \bottomrule
      \end{tabular}
    \end{table}

     \subsection{Temporal hierarchy}\label{sec:temporal_simu}
     We now consider a temporal hierarchy consisting of seven (daily) bottom-level series ($Y_1$-$Y_7$) and one (weekly) top-level series ($Y_8$).
     This setting is frequently encountered in supply chain management, where daily and weekly forecasts are required to support operational decisions (\citealp{syntetosSupplyChainForecasting2016}).
     The bottom-level time series are restricted to values of 0 or 1.
     Even with binary variables, the joint distribution of the seven bottom-level series and the total has a large number of discrete points in its domain. That makes this simulation study a good test case for the stepwise reconciliation algorithm outlined in Section \ref{sec:algorithm2}.

     \subsubsection{Simulation setup}

     While intermittent series are characterised by fluctuating demand intervals and size at lower levels, they may exhibit seasonality and trend when temporally aggregated into higher levels (\Citealp{kourentzesElucidateStructureIntermittent2021}).
     To emulate these characteristics, the data-generating process of the temporal hierarchy is based on simulating the weekly (total) time series first. This is carried out by simulating Poisson distributed data with a conditional mean that follows a seasonal AR process. We set the seasonal period and number of autoregressive terms to $4$ and $3$, respectively. The parameters are carefully tuned so that the conditional mean of the total series lies in the range $[2.5, 4.5]$. For the rare instances where simulated values of the total ($Y_8$) exceed $7$, $Y_8$ is set to $7$. All data generation is implemented with the \code{gratis} package \citep{gratis}, which ensures the diversity of the simulated series (\Citealp{kangGRATISGeneRAtingTIme2020}).
     
     Once a weekly series has been generated, these are disaggregated into daily time series. 
     Seven values are drawn from independent Beta distributions $\zeta_i\sim\textrm{B}(i, 4)$ for $i=1,\dots,7$, each corresponding to $Y_i$ for $i=1,\dots,7$.
     Letting $Y_8=w$, the values of $Y_i$ for the $w$ days corresponding to the highest values of $\zeta_i$ are set to $1$, while the remainder is set to $0$. By allowing the first parameter in the Beta distribution to increase with $i$, values of 1 are more likely to occur later in the week, implying a weekly pattern.

     For each daily time series, after discarding an initial burn-in, $903$ observations ($129$ weeks) are generated. Analogous to Section~\ref{sec:cross-sectional_simu}, we utilise the expanding window strategy with the first training window size of $350$ (50 weeks). This yields $553$ pairs of one-step-ahead forecasts and realisations (corresponding to 79 weeks). The SDFR algorithm is implemented using the first $532$ of these pairs. The final $21$ observations are used to assess the forecasting performance. This is also replicated 1000 times.

     Base forecasts of daily time series are generated utilising a logistic regression model. The covariates are the values of the previous six daily observations and weekly dummy variables. We implement the logistic regression model with the \code{glm} function in \proglang{R}. Base forecasts of weekly time series are generated using integer-valued GARCH $(\textrm{INGARCH})$ models (\Citealp{fokianosPoissonAutoregression2009}). The $\textrm{INGARCH}(p, q)$ model assumes a Poisson distribution with a conditional mean that evolves according to:
     \[
      \lambda_t = \beta_0 + \sum_{k=1}^p \beta_ky_{t-k} + \sum_{l=1}^q \alpha_l\lambda_{t-l}, 
     \] where $\beta_k, k=0,\dots,p$ and $\alpha_l, l=1,\dots,q$ are unknown parameters. $\lambda_t$ is the conditional mean at time $t$.  We fit INGARCH(3,3) into the weekly series. While methods could be used to select the correct lag order for the INGARCH, our emphasis here is on the performance of reconciliation, including under model misspecification. All implementation is carried out using the \code{tscount} package(\Citealp{liboschikTscountPackageAnalysis2017}) for \proglang{R}. The model assumes a Poisson distribution, while our domain is finite; therefore, the probability $Pr(Y_8=7)$ is set equal to $Pr(Y_8\geq 7)$ implied by the INGARCH model (in either case, this probability is small in practice).

     \subsubsection{Simulation results}
\begin{figure}
       \caption{\label{fig:sim_temporal_mcb_prob}Average ranks and 95\% confidence intervals for the four approaches in the temporal simulation. The overall ranks of the approaches in terms of Brier scores are shown to the right of their names.}
       \includegraphics[width=\textwidth]{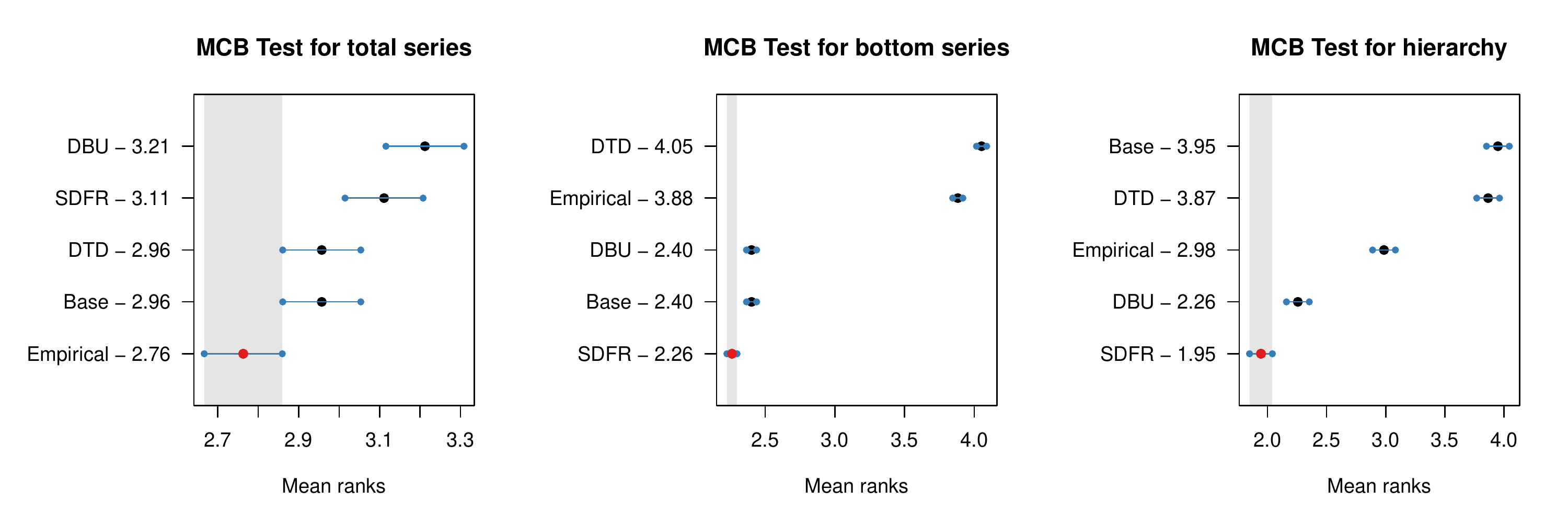}
     \end{figure}

     Table~\ref{tab:sim_temporal_res_dist} summarises the average Brier score (over 21 evaluation points and 1000 replications) of the SDFR compared to four benchmarks.
     The SDFR method outperforms the base forecast for four out of eight series and the empirical forecast for all series except the total. Most importantly, SDFR outperforms all other methods for the joint distribution.      Figure~\ref{fig:sim_temporal_mcb_prob} displays the results of MCB tests to test whether differences in forecast performance are significant. The SDFR significantly outperforms all other methods when Brier scores are computed for bottom-level series only. The same result is found for Brier scores of the joint distribution. Regarding the total $Y_8$, while SDFR performs worse than top-down and base, these differences are not statistically significant. It should, however, be noted that the empirical distribution significantly outperforms all other methods for the top-level series. This suggests that the INGARCH model is a poorly specified base model in this case. It is nonetheless encouraging that even under such misspecification, the SDFR is robust enough to produce the best forecast for the joint distribution.
     
     One intriguing observation is that the forecast accuracy of SDFR improves for $Y_4$ through $Y_7$.
     This could be attributed to the ordering used in the stepwise implementation of SDFR, which progresses from $Y_1$ to $Y_7$. The coherent forecasts of the first few series (i.e., $Y_1$, $Y_2$ and $Y_3$) are adjusted multiple times when constructing the final joint distribution, possibly introducing more noise into the forecasts of these variables. However, in contrast, the forecast accuracy of the total series does not exhibit significant deviation from base forecasts,  even though it is adjusted most frequently. This showcases the robustness of SDFR. It is crucial to note that while the results do depend on the ordering used in SDFR, this can be mitigated by implementing the algorithm with multiple orderings and then averaging the resulting coherent distributions.

     \begin{table}
     \centering
     \caption{\label{tab:sim_temporal_res_dist} Probabilistic forecast performance summary of temporal setting with Brier Score ($\times 10^{-2}$). Row-wise minimum values are displayed in \textbf{bold}.}
     \begin{tabular}{lccccc}
     \toprule
      & Base & DBU & DTD & SDFR & Empirical \\\midrule
      $Y_1$ & \textbf{40.78} & \textbf{40.78} & 49.40 & 40.95 & 49.79 \\ 
      $Y_2$ & \textbf{41.39} & \textbf{41.39} & 49.61 & 41.52 & 49.75 \\ 
      $Y_3$ & \textbf{42.06} & \textbf{42.06} & 49.86 & 42.07 & 49.73 \\ 
      $Y_4$ & 43.02 & 43.02 & 50.03 & \textbf{42.78} & 49.72 \\ 
      $Y_5$ & 43.56 & 43.56 & 50.22 & \textbf{43.09} & 49.73 \\ 
      $Y_6$ & 44.00 & 44.00 & 50.27 & \textbf{43.33} & 49.73 \\ 
      $Y_7$ & 44.31 & 44.31 & 50.28 & \textbf{43.92} & 49.78 \\ 
      $Y_8$ & 82.58 & 83.48 & 82.58 & 83.11 & \textbf{82.55} \\ 
      $Y$ & 99.54 & 97.77 & 99.36 & \textbf{97.74} & 99.12 \\ 
     \bottomrule
     \end{tabular}
     \end{table}

     \section{Empirical study}
     \label{sec:application}
     This section uses the proposed DFR algorithm to forecast two publicly available datasets.
     Section~\ref{sec:application_crime} concentrates on temporal hierarchies of the number of criminal offences in different census tracts of Washington D.C.
     Section~\ref{sec:M5} deals with a cross-sectional hierarchy using a subset of retail sales variables from the M5 dataset.

     \subsection{Application: forecasting crime in Washington D.C.}
     \label{sec:application_crime}

     Forecasting crime numbers in a specific area is vital for managing public safety and police resources. Since crime numbers are potentially serially correlated (\citealp{aldor-noimanSpatioTemporalLowCount2013}), forecasting these data should benefit from time series modelling.
     However, forecasting becomes more challenging when dealing with smaller areas with sparse crime numbers.
     The original dataset\footnote{which can be downloaded from \url{https://crimecards.dc.gov/}.} contains all reported crimes from 2014 to 2022 in Washington D.C. that have been aggregated into weekly time series. We take the data at a census tract level resulting in $231$ weekly time series.

     Our next step is to construct a temporal hierarchy for each census tract. These are two-level temporal hierarchies that consist of weekly (bottom-level) and monthly (top-level) frequencies. Strictly speaking, the data at a monthly frequency consist of a period of four weeks (28 days). However, we will use `monthly' rather than `four-weekly' for ease of exposition. We develop an individual DFR model for each census tract. While a full cross-temporal hierarchy along the lines of \cite{kourentzesCrosstemporalCoherentForecasts2019} would be an interesting exercise for this dataset, extending cross-temporal reconciliation to the discrete case lies beyond the scope of this paper.
     
     The number of crimes in the bottom-level series, is generally either 0, 1 or 2, with less than 1\% of the sample resulting in values greater than 2. Therefore, to facilitate implementation of the DFR algorithm, we therefore set the domain of the weekly series to either $[0, 1]$ or $[0, 1, 2]$, where the rare number of values above $2$ are set to $2$.

     For forecast evaluation, we generate forecasts for the month ahead, consisting of four weekly forecasts and a monthly forecast. Weekly forecasts are generated as four-step-ahead probabilistic forecasts from an INGARCH(3, 4). The monthly forecast is obtained as a one-step-ahead forecast after fitting an INGARCH(2, 2) to the data aggregated at a monthly frequency.
     A training window of 52 weeks is used to fit all models, and forecasts are evaluated on 16 weeks of data using an expanding window evaluation. We repeat the above procedure for each census tract, resulting in a total of $3696$ samples ($231$ census tracts multiplied by $16$ evaluation points per census tract).
     
\begin{figure}[h]
       \caption{\label{fig:application_crime}Average ranks and 95\% confidence intervals for the five approaches in the crime forecasting application. The overall ranks of the approaches in terms of Brier scores are shown to the right of their names.}
       \centering
       \includegraphics[width=\textwidth]{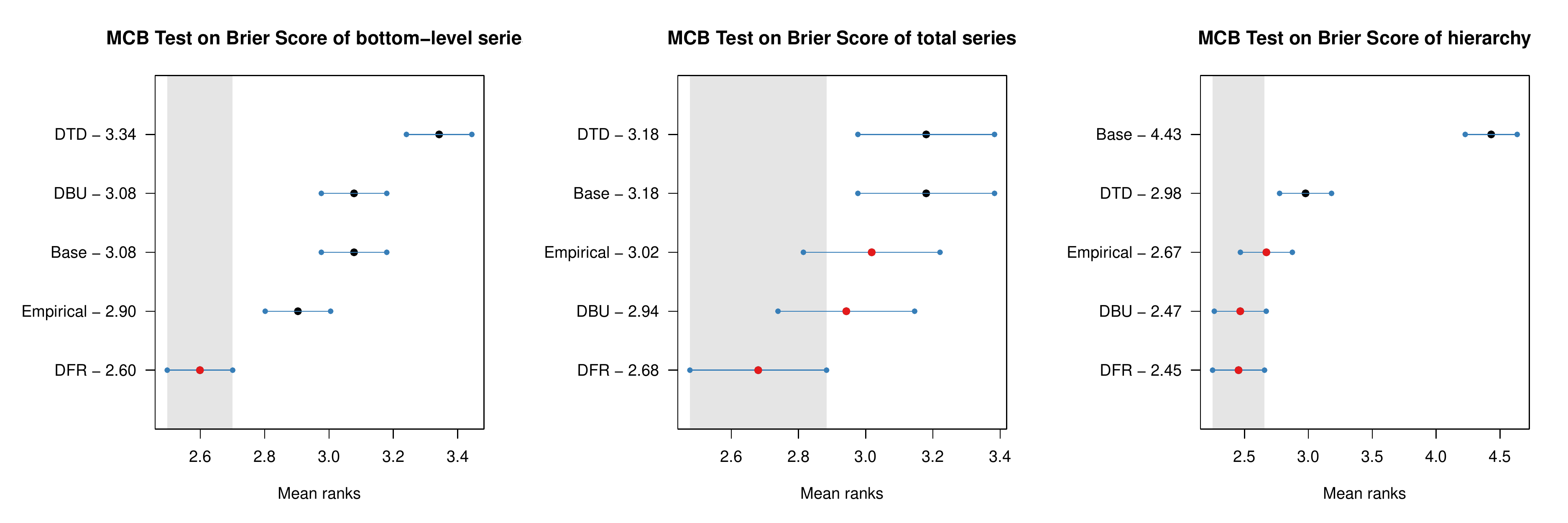}
     \end{figure}

     We calculate Brier scores at the total level, at the bottom level and for the joint distribution of the entire hierarchy.
     Table~\ref{tab:crime_bs} summarises the mean Brier scores of all methods. We also perform the MCB tests, with results shown in Figure~\ref{fig:application_crime}.

     \begin{table}[h]
       \centering
       \caption{\label{tab:crime_bs}Summarised Brier Score ($\times 10^{-2}$) of test samples in crime forecasting application. Row-wise minimum values are displayed in \textbf{bold}.}
       \begin{tabular}{cccccc}
       \toprule
       ~ & Base & DBU & DTD & DFR & Empirical \\\midrule 
       Total & 58.36 & \textbf{58.02} & 58.36 & 58.09 & 59.12 \\ 
       Bottom & 34.34 & 34.34 & 34.73 & \textbf{34.25} & 34.54 \\ 
       Hierarchy & 73.73 & \textbf{67.75} & 68.18 & {67.85} & 68.69 \\ 
       \bottomrule
       \end{tabular}
       \end{table}

     Taking these results together, DFR significantly outperforms all other methods at the bottom level. The mean Brier score for DFR over the entire hierarchy is, however, higher than the corresponding figure for the bottom-up method. In spite of this, these differences are not statistically significant, and the DFR has a lower average rank than the bottom across the 3696 evaluation points. Overall, the finding is that DFR can improve bottom-level forecasts and that improvements over the base forecasts are significant for all levels. Reconciliation methods also outperform the benchmark empirical distribution.
     
     \subsection{Application: forecasting intermittent demand in M5 dataset}
     \label{sec:M5}

     Intermittent time series are commonly observed in the fields of inventory and supply chain management, where forecasts support decisions concerning daily store replenishment, transportation plans, after-sales services and other applications (\citealp{babaiDemandForecastingSupply2022}). In this subsection, we apply the proposed SDFR algorithm to a subset of the M5 (\citealp{makridakisM5AccuracyCompetition2022}) dataset, which comprises $3049$ products sold by Walmart at ten stores located in three states in the USA from January 29, 2011 to May 23, 2016 (1941 days).
     
     We construct two-level cross-sectional hierarchies, with stores at the bottom level and states at the top level. This is done separately for individual products (stock-keeping units). We restrict our attention to low-count intermittent time series by excluding a stock-keeping unit if its maximum daily sale in any store is greater than $4$. Products with insufficient sample size (i.e., newly launched products) are also excluded. This leads to a total of $665$ hierarchies.
     
     One-day ahead forecasts are used with an expanding window, the SDFR algorithm is trained on $730$ pairs of observations and one step ahead forecasts. The last $28$ days of data are kept for evaluation. The base forecasts are produced as combinations of the SBA (\citealp{syntetosAccuracyIntermittentDemand2005}) forecasting method for intermittent demand and forecasts based on a negative binomial distribution as proposed in \cite{kolassaEvaluatingPredictiveCount2016}.

\begin{figure}[h]
      \caption{\label{fig:application_M5}Average ranks and 95\% confidence intervals for the five approaches in the M5 application. The overall ranks of the approaches in terms of average Brier scores are shown to the right of their names.}
      \centering
      \includegraphics[width=\textwidth]{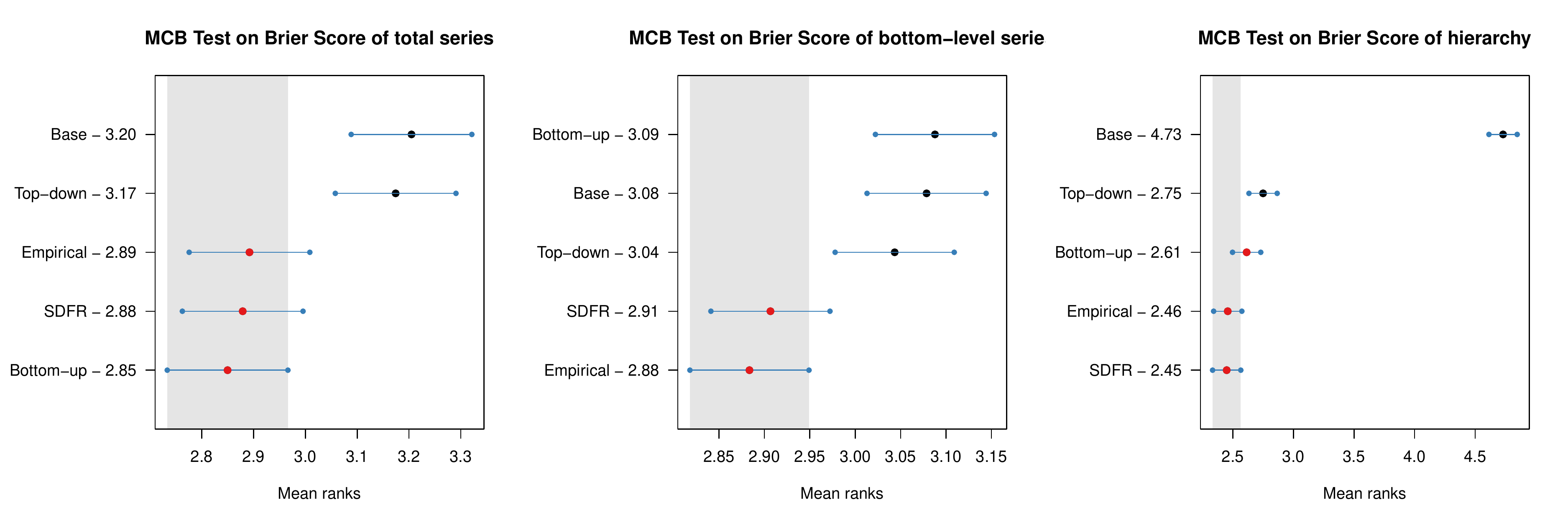}
    \end{figure}

    Table~\ref{tab:M5} presents the accuracy of benchmarks and SDFR in terms of average Brier Score across the $665$ hierarchies and $28$ evaluation periods. The corresponding MCB test is displayed in Figure~\ref{fig:application_M5}. In terms of the overall hierarchy, SDFR produces the best average Brier score. The SDFR method also significantly outperforms the discrete top-down method and base forecasts across all levels and significantly outperforms the bottom-up method when evaluation is carried out on the bottom level. The empirical benchmark performs well in this setting, perhaps due to the fact that serial dependence is weak for these data. However, the overall conclusion is that SDFR improves upon base forecasts, outperforms other reconciliation benchmarks and is competitive with the empirical benchmark.
   
    \begin{table}
        \centering
        \begin{tabular}{llllll}\toprule
            ~ & Base & DBU & DTD & SDFR & Empirical \\ \midrule
            Total & 50.86 & 50.57 & 50.86 & \textbf{50.34} & 50.35 \\ 
            Bottom & 23.97 & 23.97 & 23.90 & 23.83 & \textbf{23.82} \\ 
            Hierarchy & 62.74 & 55.87 & 56.02 & \textbf{55.63} & 55.64 \\ \bottomrule
        \end{tabular}
        \caption{\label{tab:M5}Summarised Brier Score ($\times 10^{-2}$) of test samples in M5 forecasting application. Row-wise minimum values are displayed in \textbf{bold}.}
    \end{table}

     \section{Discussion}
     \label{sec:discussion}

     Distributional forecasts are becoming increasingly important in both academic and industry settings for operational research. In the context of hierarchical forecasting, \cite{kolassaWeWantCoherent2022} suggests shifting focus from coherent point forecasts to coherent probabilistic forecasts.
     The proposed discrete reconciliation framework can generate coherent distributional forecasts for discrete-valued HTS.
     Key functional such as the median, mean, covariance, quantiles and probabilities of extreme events can be derived from the distributional forecasts.
     It is worth noting that mean point forecasts obtained by our DFR method are also coherent.

     This paper focuses on linear reconciliation, achieved by multiplying the base incoherent probability vector with a reconciliation matrix.
     To extend to non-linearity from a machine learning perspective, discrete forecast reconciliation could be viewed as a classification problem with a $q$-dimensional input mapped to an $ r$ dimensional output representing class probabilities. Machine learning models, such as specifically designed neural networks, can be used to solve this problem.
     When thousands of time series need to forecast, cross-learning techniques can be employed to construct a global non-linear reconciliation model. This poses an interesting avenue for future research.

     One issue with the proposed DFR algorithm is that its dimensionality grows both with the number of variables in the hierarchy and with the cardinality of the domain of the discrete variables. This is not so much a shortcoming of DFR, but a limitation of trying to model a high-dimensional multivariate discrete distribution. For example, even if the 665 products in the M5 application were expressed as binary variables, their multivariate distribution would have support on $666\times2^{665}$ discrete points, much greater than the number of particles in the universe. In light of this, we advocate for DFR and SDFR as methods to reconcile small to moderate-sized hierarchies. Larger hierarchies will inevitably include variables that are much better treated as continuous. Reconciling hierarchies with both discrete and continuous variables remains a challenging open question, we believe the methods we have proposed in this paper, represent an important piece in that larger puzzle.

     In order to shed more light on how the curse of dimensionality affects the computational efficiency of DFR and SDFR, we summarise the number of series, the number of parameters (elements of $\mathbf{A}$) and the estimated run-time for training the reconciliation model once in each of our experiments. 
     The results are shown in Table~\ref{tab:time}. 
     Note that we only report values of the largest hierarchy in each empirical study. 
     Computational times were evaluated on a machine equipped with a Microsoft Windows 10 system, an Intel Core i7-8700 processor clocked at 3.20GHz, providing six cores and 16 GB of memory. We suggest applying the proposed algorithms to small to moderate-sized hierarchies that have similar-size domains to those seen in our experiments. When applying DFR ad SDFR to larger hierarchies, the bottom-level series should be low-count, such as binary. 

     \begin{table}
       \centering
       \resizebox{\textwidth}{!}{
       \begin{tabular}{lcccc}
        \toprule
        & Section~\ref{sec:cross-sectional_simu} & Section~\ref{sec:temporal_simu} & Section~\ref{sec:application_crime} & Section~\ref{sec:M5}\\ \midrule
        Number of bottom-level series & 2 & 7 & 4 & 4\\
        Maximum of bottom-level series & 1 & 1 & 2 & 4\\
        Cardinality of coherent domain & 4 & 128 & 81 & 625\\
        Cardinality of incoherent domain & 12 & 1024 & 729 & 10625 \\
        Number of parameters & 22 & 1312 & 9342 & 25420 \\ 
        Algorithm & DFR & SDFR & DFR & SDFR \\
        Computational time (seconds) & 0.1 & 1.6 & 311 & 1558\\ \bottomrule
       \end{tabular}}
       \caption{\label{tab:time} Reconciliation model size and computational time (in seconds) in the four experiments implemented in different sections.}
     \end{table}

     Finally, we note that in our experiments, fairly simple models were used to obtain base forecasts. While it remains challenging to accurately forecast low count time series with an excessive number of zeros, state-of-the-art methods have been proposed by \cite{berryBayesianForecastingMany2020a} and \cite{weissEfficientAccountingEstimation2022}. While adopting such methods could lead to more accurate base and unreconciled forecasts, our emphasis here was to demonstrate that forecast reconciliation can improve base forecasts in the discrete case.

     \section{Conclusion}
     \label{sec:conclusion}

     This paper develops a novel forecast reconciliation framework for count hierarchical time series.
     While our proposed idea of mapping probabilities from incoherent points to coherent points, is analogous to methods for continuous data, implementation details vary significantly.
     Our novel framework involves rewriting the optimisation of the Brier score as an assignment problem facilitating a solution using quadratic programming. 
     To address computational concerns in higher dimensions, we introduce a stepwise discrete reconciliation algorithm by breaking down a large hierarchy into smaller ones.
     We also propose probabilistic extensions of traditional top-down and bottom-up methods for count time series.

     Our DFR and SDFR algorithms produce coherent probabilistic forecasts and improve forecast accuracy.
     We demonstrate this through simulation experiments on cross-sectional and temporal hierarchies, where our algorithms outperform discrete top-down and discrete bottom-up approaches as well as a benchmark based on the empirical distribution. Additionally, we apply the DFR algorithm to forecast crime numbers in Washington D.C. and sales of products in the M5 dataset with promising results.
     
     One key factor contributing to our strong results is the utilisation of forecast combinations.
     Similar to reconciliation approaches for continuous variables, our framework combines forecasts and the information used to produce forecasts at different levels.
     It is also important that our models train the reconciliation weights using out-of-sample forecasts generated by the expanding window strategy, leading to robust results.

     While this work provides an important step in hierarchical time series forecasting, future research should focus on hierarchies with both continuous data and count data. In light of this, count hierarchical time series forecasting remains an open issue that requires further attention from researchers in the future.

\section*{Acknowledgements}

Yanfei Kang is supported by the National Natural Science Foundation of China (No. 72171011). This research was supported by international joint doctoral
education fund of Beihang University and the high-performance computing (HPC) resources at Beihang University.

\newpage

\appendix

\section{Algorithms}
\label{appendix:adjust}

The \code{Adjust} algorithm is used to adjust an existing multivariate joint distribution to make its marginalisation over one of the variables equal to that given by a reconciled distribution in a single step of the stepwise procedure.

\begin{algorithm}[H]
  \label{alg:adjust}
  \caption{\code{Adjust}}
  \SetKwInOut{Input}{Input}
  \SetKwInOut{Output}{Output}
  \Input{$\bpi(y_0,y_1,\dots,y_i), \tilde\pi_i, y_i \in \{0,1,\dots,k_i\}$}

  $\bpi(y_0,\dots,y_{i-1}) = \sum_{y_i}\bpi(y_0,\dots,y_i)$\;
  $\pi_i = \sum_{y_0,\dots,y_{i-1}}\bpi(y_0,\dots,y_i)$ \;
  \For { $j = 0,\dots,k_i$} {
    $\bpi'(y_0,\dots,y_{i-1}, y_i=j) = \bpi(y_0,\dots,y_{i-1}) \times \frac{\tilde\pi_i}{\pi_i}$ \;
  }

  \Output{$\bpi'(y_0,\dots,y_i)$}

 \end{algorithm}

 The \code{ConstructJointDist} algorithm constructs a new joint distribution from two other joint distributions.
 Consider variables $y_1, ~ \bY_2, ~ y_3, ~ y_4, ~ \bY_5$, where $\bY_2$ and $\bY_5$ are vectors and $y_1, ~ y_3, ~ y_4$ are scalars.
 They have the following relations.
 \[
  y_1 = |\bY_2|_1 + y_3, \quad y_3 = y_4 + |\bY_5|_1,
 \]
 where $|\cdot|_1$ represents the sum of all the variables in the vector.
 The given distributions are $\bpi(y_1, ~ \bY_2, ~ y_3)$ and $\bpi(y_3, ~ y_4, ~ \bY_5)$.
 The marginal distributions of $y_3$ derived from the two distributions are the same, i.e.,
 \[
  \sum_{y_3} \bpi(y_1, ~ \bY_2, ~ y_3) = \sum_{y_3}\bpi(y_3, ~ y_4, ~ \bY_5)
\]
 Assuming the joint distribution of $y_1$ and $\bY_2$ is independent of the joint distribution of $y_4$ and $\bY_5$ given $y_3$, we can obtain the probability of one point in the new joint distribution using the following equation: \[
   \begin{aligned}
  &\text{Pr}(y_1=a_1, ~ \bY_2=\mathbf{a}_2, ~ y_4=a_4, \bY_5 = \mathbf{a}_5) =\\ &\text{Pr} (y_1=a_1, ~ \bY_2=\mathbf{a}_2, ~ y_3=a_3) \times \text{Pr}(y_4=y_4, ~ \bY_5=y_5|y_3=a_3).
   \end{aligned}
 \]
 The equation constructs the joint distribution of $y_1, \bY_2, y_4, \bY_5$ by eliminating the shared variable $y_3$ of the two distributions.

\newpage
\begingroup
\setstretch{1.15}
\bibliographystyle{agsm}
\bibliography{references.bib}
\endgroup

\end{document}